# Data stream mining for predicting software build outcomes using source code metrics

**Dr Jacqui Finlay, Dr Russel Pears & Dr Andy M. Connor**

*Auckland University of Technology, Auckland, New Zealand*

## ABSTRACT

*Context:* Software development projects involve the use of a wide range of tools to produce a software artifact. Software repositories such as source control systems have become a focus for emergent research because they are a source of rich information regarding software development projects. The mining of such repositories is becoming increasingly common with a view to gaining a deeper understanding of the development process.

*Objective:* This paper explores the concepts of representing a software development project as a process that results in the creation of a data stream. It also describes the extraction of metrics from the Jazz repository and the application of data stream mining techniques to identify useful metrics for predicting build success or failure.

*Method:* This research is a systematic study using the Hoeffding Tree classification method used in conjunction with the Adaptive Sliding Window (ADWIN) method for detecting concept drift by applying the Massive Online Analysis (MOA) tool.

*Results:* The results indicate that only a relatively small number of the available measures considered have any significance for predicting the outcome of a build over time. These significant measures are identified and the implication of the results discussed, particularly the relative difficulty of being able to predict failed builds. The Hoeffding Tree approach is shown to produce a more stable and robust model than traditional data mining approaches.

*Conclusion:* Overall prediction accuracies of 75% have been achieved through the use of the Hoeffding Tree classification method. Despite this high overall accuracy, there is greater difficulty in predicting failure than success. The emergence of a stable classification tree is limited by the lack of data but overall the approach shows promise in terms of informing software development activities in order to minimize the chance of failure.

## 1. Introduction

Software development projects involve the use of a wide range of tools to produce a software artifact, and as a result the history of any given software development endeavor may be distributed across a number of such tools. Recent research in this area [1] has classified the types of artifacts that can be used to reconstruct the history of a project. These include the source code itself, source code management systems, issue tracking systems, messages between developers and users, meta-data about the projects and usage data. Software repositories such as source code management systems have become a focus for emergent research as being a potential source of rich information regarding software development projects. The mining of such repositories is becoming increasingly common with a view to gaining a deeper understanding of the development process.

Jazz (http://jazz.net) is a collaborative software engineering toolset developed by IBM that has been recognized as offering new opportunities in this area because it integrates the software archive and bug database into a single repository. This is achieved by linking bug reports and source code changes with each other [2]. In addition to the toolset itself, IBM has also released the data repository associated with the development of Jazz. This provides access to nearly all of the artifacts that can be used to construct the history of the project [1] and includes developer communication as well as source code and bug reports. Such a repository provides much potential in gaining valuable insights into the development process yet comes with particular challenges. One of the main challenges is that the Jazz environment is not designed to maintain a complete history of build events. The implication of this challenge is that mining the repository for a given project does not involve a static base of data which grows over time. Instead, the size of the data is essentially fixed but the bounds of the data change over time meaning that the available data at any instant is just a snapshot of the total development history.

Traditional data mining methods and software measurement studies are tailored to static data environments. These methods are typically not suitable for streaming data which is a feature of many real-world applications. Software project data is produced continuously and is accumulated over long periods of time for large systems. The dynamic nature of software and the resulting changes in software development strategies over time causes changes in the patterns that govern software project outcomes. This phenomenon has been recognized in many other domains and is referred to as data evolution, dynamic streaming or concept drifting [3]. However there has been little research to date that investigates concept drifting in software development data despite being recognized as an area of interest [4]. Changes in a data stream can evolve slowly or quickly and the rates of change can be queried within stream-based tools. This paper describes an initial attempt to fully extract the richness available in the Jazz data set by constructing predictive models to classify a given build as being either successful or not, using software metrics as the predictors for a build outcome. In this context, a build attempt in the Jazz software is the process of performing software subsystem integration and the integration of multiple subsystems. Build success is therefore a successful integration which includes compilation, testing and packaging without producing any errors [5]. Build failure is therefore the presence of an error that prevents the creation of a deployable software package.

This research investigates the construction of predictive models to determine build outcomes in advance of the build attempt. Previous work [6, 7] in this area has shown that there is potential for such prediction to occur, however the use of traditional data mining approaches results in unstable models with limited applicability. This work therefore investigates the application of data stream mining approaches that are capable of adapting to changes in the underlying data distribution to determine whether they produce more stable models. The goals of the research are formally expressed in section 4.

The models that are built in this research involve the tracking of the trajectory of classification accuracy over time. The models are built incrementally as each new data instance (software build) that arrives is used to update the existing model. With the Jazz repository there exists a certain set of builds having known outcomes that can be chosen from to train and induce an initial model (in the form of a Decision Tree) and the rest of the builds can then be utilized to incrementally update the model. This implies that there is freedom to choose in which order the instances are supplied and hence this produces an opportunity to track the effects of instance ordering on classification accuracy of the predictive model that are constructed and incrementally maintained.

Section 2 provides a brief overview of related work. Section 3 discusses the nature of the Jazz data repository and software metrics that were utilized during the mining the repository. Section 4 discusses the approach to mining the software repository in Jazz, and initial results are presented in section 5. Finally, in sections 6 and 7, a discussion of the limitations of the current work and a plan for addressing these issues in future work are presented.

## 2. Background & related work

The mining of software repositories involves the extraction of both basic and value-added information from existing software repositories [8]. Such repositories are generally mined to extract facts by different stakeholders for different purposes. Data mining is becoming more prevalent in software engineering environments [4, 9, 10] and the application of mining software repositories include areas as diverse as the development of fault prediction models [11], impact analysis [12], effort prediction [13, 14], similarity analysis [15] and the prediction of architectural change [16] to name but a few. The growth in popularity in mining software repositories have led some researchers to believe that we are on the brink of introducing the idea of Software Intelligence (SI) as the future of mining software engineering data [17]. Hassan and Xie [17] argue that "Software Intelligence offers software practitioners (not just developers) up-to-date and pertinent information to support their daily decision-making processes. SI provides practitioners with access to specialized fact-supported views of their software system so they can answer critical questions about it."

According to Herzig & Zeller [2], Jazz offers not only huge opportunities for software repository mining but also a number of challenges. One of the opportunities is the provision of a detailed dataset in which all software artifacts are linked to each other. To date, much of the work that utilizes Jazz as a repository has focused on the convenience provided by linking artifacts such as bug reports to specification items along with the team communication history. Researchers have focused on areas such as whether there is an association between team communication and build failure [5] or software quality in general [18]. Other work has focused on whether it is possible to identify relationships among requirements, people and software defects [19] or has focused purely on the collaborative nature of software development [20]. To date, most of the work involving the Jazz dataset has focused on aspects other than analysis of the source code contained in the repository and as such the full range of the available project history is not being fully utilised.

Research that focuses on the analysis of metrics derived from source code analysis to predict software defects has generally shown that there is no single code or churn metric capable of predicting failures [21-23]. However, evidence suggests that a combination of such metrics can be used effectively [24]. To date no such source code analysis in a data stream context has been conducted on the Jazz project data and it is the objective of this study to perform an in-depth analysis of the repository to gain insight into the usefulness of software metrics in predicting software build failure. In particular the focus is on using data stream mining techniques to enable software development teams to collect and mine SE data on the fly to provide rapid just-in-time information to guide software decisions as has been suggested in the literature [4]. However, this is just one element of a larger research agenda that looks to fully integrate the available project history into a usable decision support environment that supports the software development team. It is likely that such an approach will use not only software metrics, but also developer communication metrics [5] as well as repository level change measures. Such measures have been noted as being good indicators of failure in the literature [25].

Despite the claims that change measures are good indicators of failure, there is also evidence that fine grained source code change metrics are also good indicators of failure when compared to code churn metrics [26]. The conflicting evidence of what is likely to be a good indicator of failure implies that there is a need for a decision support dashboard that is based on multiple measures. The need for decision support in software engineering has been identified in the literature [27] and is clearly concordant with the concepts of Software Analytics [28] and Software Intelligence. The emergence of decision support models based around the use of software metrics is likely to see different aspects of software engineering research merge and overlap. This is already apparent with recent research that looks at using Search Based Software Engineering techniques as a feature selection technique applied to choosing software metrics for defect prediction [29]. Other approaches investigate the use of information density as a means of selecting between metrics [30]. Despite this potential for overlap much of the recent research on software metrics has focused on some form of static analysis for determining fault proneness or defect rates [31].

Much of the body of knowledge on software metrics relates to the use of measures at a low level of abstraction that results from the simple fact that metrics are calculated at the micro level (e.g. method, class, package). Vasilescu, Serebrenik and van den Brand [32] argue that metrics should therefore be aggregated at the macro-level (e.g. system) in order to provide useful measures that provide insight in terms of the study of maintainability and evolution of the system as a whole. Buse and Zimmerman [28] also suggest that whilst software projects can be characterised by a range of metrics that describe the complexity, maintainability, readability, failure propensity and many other important aspects of software development process health, it still continues to be risky and unpredictable. In their paradigm of software analytics, Buse and Zimmerman also suggest that metrics themselves need to be transformed into measures that can be utilised to gain insights. As such it is necessary to distinguish questions of information which some tools already provide (e.g., how many bugs are in the bug database?) from questions of insight which provide managers with an understanding of a project's dynamics (e.g., will the project be delayed?). They continue by suggesting that the primary goal of software analytics is to help managers move beyond information and toward insight, though this requires knowledge of the domain coupled with the ability to identify patterns involving multiple indicators. There are many parallels between the Software Analytics paradigm and the concept of Software Intelligence proposed by Hassan and Xie [17].

In previous work [6, 33, 34] source code analysis has been conducted using software metrics derived from the Jazz project by applying traditional (static) data mining methods. This was conducted to gain insights into predicting build failure that could be potentially useful in terms of determining the likely outcome of a given build. This previous work is completely superseded by the work presented in this paper as the use of a data stream mining techniques are demonstrated to be more robust and reliable. It is our belief that the Jazz dataset provides sufficiently rich information to extend the usefulness of such models by developing a dynamic analysis tool that integrates with and supports the decision making process throughout an entire build cycle and encapsulates the goals of both the Software Analytics paradigm and supports the tenets of Software Intelligence, particularly in the area of attempt to complement practitioner expertise rather than attempting to replace it.

## 3. The Jazz dataset

### 3.1 Overview of Jazz

IBM Jazz is a fully integrated software development tool that automatically captures software development processes and artifacts. The Jazz repository contains real-time evidence that allows researchers to gain insights into team collaboration and development activities within software engineering projects [35]. With Jazz it is possible to extract the interactions between contributors in a development project and examine the artifacts produced. This means that Jazz provides the capability to extract social network data and relate such data to the software project outcomes [5]. Figure 1 illustrates that through the use of Jazz it is possible to visualize members, work items and project team areas.

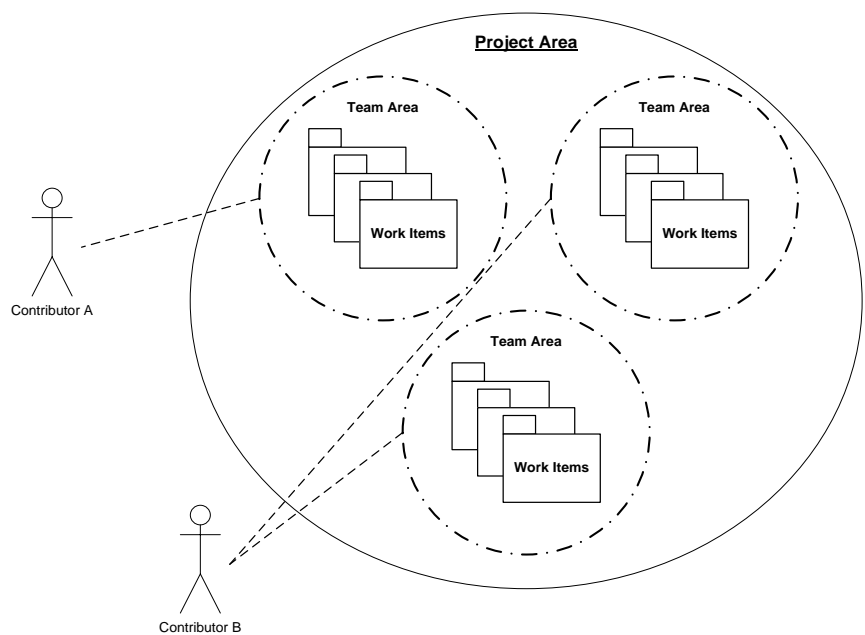

**Figure 1. Jazz Repository: Contributors, Project Area, Team Areas and Work Items.**

The Jazz repository artifacts include work items, build items, change sets, source code files, authors and comments. This provides significant opportunity to extract more data from the repository to support decisions related to the development effort. A work item is a description of a unit of work, which is categorized as a task, enhancement or defect. A build item is compiled software that forms a working unit. A change set is a collection of code changes in a number of files. In Jazz a change set is created by one author only and relates to one work item. A single work item may contain many change sets. Source code files are included in change sets and over time can be related to multiple change sets. Authors are contributors to the Jazz project. Comments are recorded communication between contributors of a work item. Comments on work items are the primary method of information transfer among developers in a global virtual team.

There are limitations for incorporating the Jazz repository into research. Firstly, the repository is highly complex and has huge storage requirements for tracking software artifacts. This is in part one of the reasons why the repository does not store a complete history. Another issue is that the repository contains holes and misleading elements which cannot be removed or identified easily. This is because the Jazz environment has been used within the development of itself; therefore many features provided by Jazz were not implemented at early stages of the project which results in challenges as to how to deal with such inconsistency. The proposed approach delves further down the artifact chain than most previous work using Jazz to access the source code as a means of bypassing holes and misleading data in artifacts that exist at higher levels of abstraction. It is the premise of this study that the early software releases were functional, so whilst the project "meta-data" may be missing details (such as developer comments) the source code should represent a stable system that can be analyzed to gain insight regarding the development project.

*3.2 Software Metrics*

Software metrics have been utilised in order to deal with the sparseness of the data that exists at higher levels of abstraction in the repository, such as developer comments. Software metric values are determined by extracting source code from the Jazz repository. Software metrics are commonly used within model-based project management methods to measure various attributes of the project, in particularly they are used to measure the complexity, quality and effort of a software development project [36]. The Jazz database contains over 200 relations, containing numerous cryptic fields. Thus data extraction via SQL queries runs the high risk of retrieving unreliable or incomplete data. Instead, this study uses the Jazz client/server APIs, an approach recommended in a study by Nguyen, Schröter, and Damian [35].

The Jazz repository consists of various types of software builds. Included in this study were continuous builds (regular user builds), nightly builds (incorporating changes from the local site) and integration builds (integrating components from remote sites). Source code files were extracted for each available build within the repository. Subsequently software metrics for each file or package were calculated by utilizing the IBM Rational Software Analyzer tool (RSA). The full range of supported metrics were calculated according to the categories described in the RSA documentation as basic metrics, dependency metrics, complexity metrics, cohesion metrics and Halstead metrics. In total, 42 different measures were calculated which are listed in Table 1.

| Complexity Metrics | Halstead Metrics | Basic Metrics |
|---|---|---|
| Average block depth | Number of operands | Depth of Inheritance |
| Weighted methods per class | Number of operators | Number of attributes |
| Maintainability index | Number of unique operands | Average number of attributes per class |
| Cyclomatic complexity | Number of unique operators | Average number of constructors per class |
| | Number of delivered bugs | Average number of comments |
| **Dependency Metrics** | Difficulty level | Average lines of code per method |
| Abstractness | Effort to implement | Average number of methods |
| Afferent coupling | Time to implement | Average number of parameters |
| Efferent coupling | Program length | Number of types per package |
| Instability | Program level | Comment/Code Ratio |
| Normalized Distance | Program vocabulary size | Number of constructors |
| | Program volume | Number of import statements |
| **Cohesion Metrics** | | Number of interfaces |
| Lack of cohesion 1 | | Lines of code |
| Lack of cohesion 2 | | Number of comments |
| Lack of cohesion 3 | | Number of methods |
| | | Number of parameters |
| | | Number of lines |

**Table 1. Software Metrics**

The study was limited to these metrics as they would be readily available to the IBM Jazz development team as a result of the tight integration between components in the Rationale software suite, particular Rational Team Concert and Rational Software Analyzer. Previous work [7] involved conducting a systematic set of experiments to determine the significance of the metrics in terms of predicting build outcomes and potentially reduce the number of metrics considered. This work applied different feature selection techniques across a range of different traditional classifier methods and one of the observations was that feature selection approaches offered limited improvement over no feature selection. As a result, in this current study all measures are used with no attempt at reducing the size of the feature space.

*3.3 Dataset Interpretation*

In the Jazz dataset a given build consists of a number of different work items. Each software build contains *changesets* that indicate the actual source code files that are modified during the implementation of the build. This is shown diagrammatically in Figure 2. The work item is the main decomposition element in the Jazz environment. A work item is essentially a development task of some form, typically an enhancement of a bug fix. Each work item will include a number of source code files. The before state of a work item can therefore be considered the contents of the source code files prior to the task being undertaken. Once the task has been undertaken the changed source code files represent the after state of the work item.

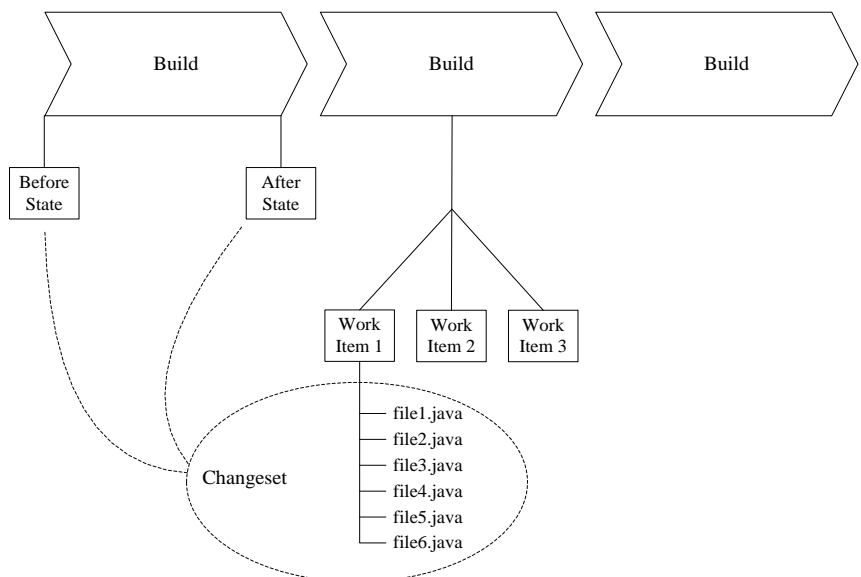

**Figure 2. Jazz Repository: Contributors, Project Area, Team Areas and Work Items.**

A given work item also includes other files such as ancillary documentation. For the purpose of this research the collection of source code files in a build that excludes these ancillary files is referred to as a changeset. Source code file membership of a changeset is not exclusive, so a given source code file may appear in multiple work items in the same build. A build typically consists of the integration of multiple work. At the lowest level of detail, a build attempt in the Jazz software is therefore the process of performing software subsystem integration and the integration of multiple subsystems. Build success is therefore a successful integration which includes compilation, testing and packaging all performed without producing any errors [5].

For this research, source code metrics for each source code file included in a build are calculated using the IBM Software Analyzer tool. The *after* state was utilized in order to ensure that the source code snapshot represented the actual software artifact that either failed or succeeded.

One challenge with this work is to consider how to represent the range of metrics values present in the different source code files and aggregate this into a suitable value for artifacts that exist at a different level of abstraction, namely the work items and builds in the Jazz environment. A range of traditional aggregation approaches have been considered, for example using the maximum value, the mean, the median or the summation of metrics. These have been compared to the proprietary aggregation algorithm contained in the IBM Software Analyzer tool and found to be less effective. It has been noted in the literature that traditional aggregation methods are generally not effective because metrics data is often skewed the interpretation of such measures becomes unreliable [32]. Another challenge with the current work is that it has been acknowledged in the literature that one of the possible causes of poor uptake on metrics-based analysis is that threshold values for certain metrics are not clearly understood [37].

## 4. Research design

**This research investigates the construction of predictive models to determine build outcomes in advance of a software build attempt by considering the application of data stream mining approaches. The overall design of the research is based on the Knowledge Discovery from Databases (KDD) process defined by Fayyad, Piatetsky-Shapiro & Smyth [38]. This process is illustrated in**

Figure 3.

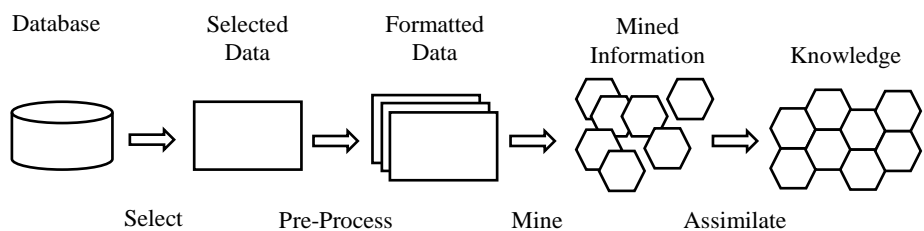

**Figure 3. KDD Process (Adapted from Fayyad, Piatetsky-Shapiro & Smyth [38])**

**As shown in**

Figure 3, a typical data mining process has four main steps. The first step is to select the types of data that will be used by the mining algorithm. This typically involves selecting relevant data from multiple sources and producing a single repository. In the case of this research, this step has been significantly simplified by the use of the Jazz data and consisted of the selection of particular builds by excluding builds for which there was no source code. The second step is to pre-process the data for analysis. Usually the data has to be formatted, sampled, adapted, and sometimes transformed for the data mining algorithm. Formatting usually involves treating noisy or missing data. Adaptation is frequently necessary to make the data work with the data mining technique that will be used. For example, some algorithms only work with nominal data. Again, the selection of the numeric valued software metrics has simplified this step as the classification tree algorithms selected work effectively with numeric data. After pre-processing, the data is ready to be mined using a data mining algorithm. The data mining step aims to extract interesting information from this data. This step may involve very different data mining techniques. In this research, the mined information is contained in the derived model itself. The last step of the data mining

process is to assimilate the mined information. This is done by interpreting and assimilating the information the data mining technique considered "interesting." In the case of model building, this step consists of evaluating the robustness and effectiveness of the produced models by using techniques such as re-substitution and cross-validation. If accepted, the model can then be incorporated into the organization's development and decision making processes.

The research aims can be expressed in more detail using the Goal-Question-Metric approach [39] which has also been used in previous research related to mining data repositories [40]. Based on this approach, the following are defined:

**Goal:** Identify stable models for build success prediction
**Question:** Do data stream mining approaches facilitate more stable prediction models than traditional data mining approaches?
**Metric:** The degree of similarity between decision tree models produced at different time periods.

The degree of similarity is evaluated in a number of ways which includes a comparison of the characteristics of the decision trees in terms of the depth of the tree, the number of decision nodes and so forth. We also define an objective measure called attribute churn, which is a measure of the degree of change in the attributes used in subsequent instances of the decision tree. We propose a normalized measure that is calculated by:

$$Attribute\ Churn = \frac{N_A + N_D}{N_T} \times 100$$

Where $N_A$ is the number of attributes added to the decision tree that were not present in the previous instance, $N_D$ is the number of attributes deleted from the previous decision tree instance so are no longer used and $N_T$ is the total number of attributes in the current decision tree.

Whilst the overall goal of this research is to determine whether data stream mining techniques produce more stable models than traditional data mining approaches, it is also important to consider the accuracy of the models as well. As such, this research consists of two phases of experimentation. The first phase of experimentation maps directly to the "mine" and "assimilate" phases of the KDD process shown in Figure 3, where the data mining is conducted to determine a build prediction model using the Hoeffding tree [41] approach and a validation of the model is conducted by varying the order in which the data is provided to the classifier. In the second phase, the models from the Hoeffding tree approach are compared with models developed using the j48 classifier. This comparison investigated the attribute churn between subsequent models produced by each method to enable the stability to be determined. This is the primary means of testing whether the goal of the research has been achieved.

### 4.1 Experimental method

Data mining is a knowledge discovery process that involves a sequence of tasks starting from understanding the problem domain and available data, through data preprocessing to deployment of the results [42]. One of the key aspects of data mining is the selection of an appropriate algorithm from the many that are available [43]. This work revolves around the use of data stream mining techniques for the analysis of software metrics. For this purpose MOA, or the Massive Online Analysis, software environment was used [44]. Data streams provide unique opportunities as software development dynamics can be examined and captured through the incremental construction of models over time that predict project outcome. Two outcomes are possible: success, or failure. A successful outcome signals that each of the constituent work items in a project has been built as per the specification and that the items have been integrated with each other without generating any errors. On the other hand, the failure outcome is caused by one or more work items malfunctioning and/or errors being generated in the integration process.

In terms of the data, each instance contains a range of complexity, dependency, Halstead, cohesion and basic software metrics. In total there are 42 measures calculated using IBM Rational Software Analyzer tool, plus the class label associated with each instance (project), as extracted from the Jazz repository. The metric values at individual file level were aggregated into a representative measure at the build level using the functionality available in the IBM Rational Software Analyzer tool.

The core data mining method used to build the predictive model is the Hoeffding tree [41] which is widely used in data stream mining. The implementation of the Hoeffding tree available from the MOA environment was used in this study. The Hoeffding tree method requires a certain number of training instances for inducing an initial classification model and in this work 20 instances were for this initial training. After the initial Hoeffding tree was induced each new instance that arrived is added to the training pool and an incremental update to the model is performed to reflect the contribution of the newly arrived instance. The updated model is then used to classify the new instance. Since the class labels of every instance was known in advance, the classification accuracy could thus be updated on an instance by instance basis. Thus in effect, training and testing are interleaved activities that typically take place in a data stream environment such as MOA.

Two types of experiments were conducted. Firstly all the instances were ordered in chronological order and the first 20 instances were used to induce an initial decision tree. The remainder of the instances were then used to incrementally train and update the initial model. The classification accuracy and other performance metrics were tracked at certain key ranges in order to gain an understanding of performance trajectory with the size of the training set used. Based on initial experimentation the training set was divided into 4 contiguous ranges, which are referred to as *phases*. In total four phases were used with phase boundaries defined as: Phase 1, covering instances from 1 to 40, Phase 2 spanning 41 to 80, Phase 3 covering instances from 81 to 180 and Phase 4 covering the rest of the available instances up to the 198$^{th}$ instance mark.

In our second experiment the order in which instances were fed to the Hoeffding tree was varied. This was done as follows. After isolating the initial pool of 20 instances which were used to construct the initial model the remaining pool of N instances was considered to be available. This pool of N instances in chronological order were then segmented into 10 groups G1, G2, …, G10. These 10 groupings were then used to define 10 different sequence orderings in the following manner. The first sequence was defined by G10, G1, G2, …, G9; the second by G9, G1, G2, …,G8, G10; the third by G8, G1, G2, …, G7, G9, G10; and so on. The tenth sequence of G1, G2, …, G9, G10 preserves pure chronological ordering. The effects of varying instance order by tracking overall classification accuracy and other metrics during the final stage of model construction were evaluated – i.e in the interval spanning the arrival of the last 21 instances. This option is preferred instead of tracking at earlier points as in experiment 1 as the model generated during this final phase would have matured to such an extent so as to enable us to test the effects of instance ordering with greater confidence than with the earlier formative stages of model development.

Various challenges need to be overcome in mining real world data. Real world data in its raw form is not always suitable for the mining process. There is often noise within the data, missing data, or even misleading data that can have negative impacts on the mining and learning process [45]. The project data that is extracted from Jazz was gathered during the development of Jazz. As a consequence, features that automatically capture project processes did not exist until later development stages of Jazz (gaps would often appear at early stages of the project data set). The methods used to mine the simulated data stream are described in the following section.

*4.1.1 Hoeffding Tree*

Decision trees were selected as the machine learning outcome for this research. This choice was influenced by the fact that the decision tree has proven to be amongst the most accurate of machine learning algorithms while providing models with a high degree of interpretability [46]. Decision trees are used to classify instances by sorting them based on feature values and such classification approaches are typically strong in modeling interactions [47] making them very suited to this study. Traditional decision tree based classifiers such as J48 [48, 49] (a Java implementation of the C4.5 algorithms) have been identified as one of the most popular data mining techniques [50]. However, these algorithms cannot easily be deployed in a data stream environment as they are incapable of incrementally updating the model, a key requirement in a data stream environment. Therefore an incremental version called the Hoeffding tree was deployed in this study.

The Hoeffding tree is a commonly used incremental decision tree learner designed to operate in a data stream environment [51]. Unlike in a static data environment decision tree, learners in a data stream environment are faced with the difficult choice of deciding whether a given decision node in the tree should be split or not. In making this decision, the information gain measure that drives this decision needs to take into account not just the instances accumulated in the node so far, but must also make an assessment of information gain that would result from future, unseen instances that could navigate down to the node in question from the root node. This issue is resolved through the use of the Hoeffding bound [52].

The Hoeffding bound is expressed as:

$$\epsilon = \sqrt{\frac{R^2 \ln(\frac{1}{\delta})}{2n}}$$

Where R is a random variable of interest, n is the observations and $\delta$ is a confidence parameter. Essentially, the Hoeffding bound states that with confidence $1 - \delta$, the population mean of R is at least $\bar{r} - \epsilon$, where $\bar{r}$ is the sample mean computed from n observations.

In the context of data mining, the variable R is the information gain measure which ranges from 0 to 1 in value. One key advantage of the bound is that it holds true irrespective of the underlying data distribution, thus making it more applicable than bounds that assume a certain distribution, for example the Gaussian distribution.

The Hoeffding tree implementation available from MOA was used in this research and coupled with a concept drift mechanism called the Adaptive Sliding Window (ADWIN). Since entire data streams are generally not stored, only a subset of the stream is available at any given time. The subset of a stream is composed of discontinuous elements which form sub-streams. These sub-streams are used by window models, where a window is determined from a particular time period of a given data stream. There are many different window models; each of which is based on different window sizes, update intervals and window closure constraints. Windows size can be time based or count based (fixed number of elements). Windows update strategies govern the expiry of older elements of data upon the arrival of new data.

Many machine learner algorithms in a data stream environment use fixed size sliding windows to incrementally maintain models. The sliding window approach is best described by considering the data stream to be fixed and the window defined by its two end points. When the window slides, each endpoint is moved some degree relative to the datastream. When the window size is fixed, the endpoints both slide along the datastream the same amount. This is shown in Figure 3 where for illustration purposes the windows slides just a single data element in the stream, which is known as eager updating of the window. In practice a sliding window may update on the arrival of more than one data element, which is known a lazy updating. When the updating interval exceeds the size of the window itself, the windows are considered to be jumping windows rather than sliding windows. Describing the sliding window approach relative to affixed datastream clarifies why the approach is called a sliding window, though in reality it is the creation of new data that creates the slide by pushing data through the window.

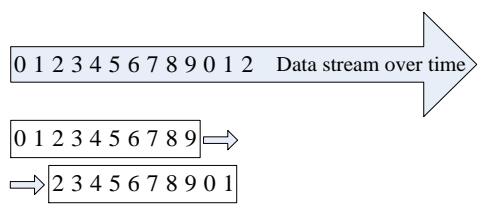

**Figure 4. Fixed Sliding Window**

Sliding windows offer the advantage of purging old data while retaining a manageable amount of recent data to update models. However, whilst the concept of windows is attractive because memory requirements are kept within manageable bounds, fixed sized window are often sub-optimal from the viewpoint of model accuracy. This is due to the fact that concept change or drift may not align with window boundaries. When changes occur within a window, all instance before the change point should be purged leaving the rest of the instances intact for updating the model built so far. The ADWIN approach has many merits in terms of detecting concept drifts in dynamic data streams.

*4.1.2  Concept Drift Detection*
ADWIN is a parameter-free adaptive sliding window strategy that compares all adjacent sub-windows to a partition window that contains all the data [53]. This method is recognised to generate the best accuracy, however may have a time cost with larger streams. This method dynamically adjusts the size of a window and derives efficient variations using Hoeffding's bound. A window will become larger when the data is stationary to maintain better accuracy. A window will become smaller when change is taking place as it will discard stale data. This eliminates the need for the user to determine/estimate what the best window size is. ADWIN works on the principle of statistical hypothesis testing. It maintains a window consisting of all instances that have arrived since the last detected change point, represented as $W_1$ in Figure 5.

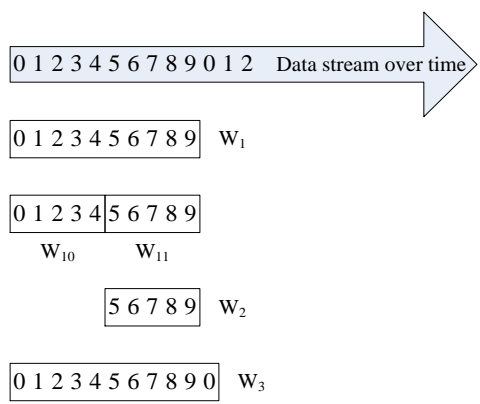

**Figure 5. Adaptive Sliding Window**

In its most basic form, the arrival of each new instance causes ADWIN to split the current window into all possible combinations of sub-windows, for example, ($W_{10}$) and ($W_{11}$). The sample means of the data in the two sub-windows are compared under a null hypothesis H0 that the means across the sub-windows are not significantly different from each other. If H0 is rejected, concept drift is taken to have occurred and ADWIN shrinks the window to only include instances in the right sub-window ($W_2$), thus removing instances in the left window representing the "old" concept. Simultaneously, the Hoeffding tree is updated to remove sub-tree(s) representing the old or outdated concept. On the other hand, if H0 is accepted then the size of the window is increased by adding the next instance in the data stream, as shown by $W_3$. In this case, the adaptive sliding window has the same effect as implementing an alternative to sliding windows, known as a landmark window. In the landmark window approach one of the endpoints of the window maintains a fixed position in the datastream and the windows is allowed to grow rather than slide. The ability to be responsive to the data and produce the same outcomes as other windowing strategies is one of the key benefits of the adaptive sliding window approach.

## 5. Results

The experimental approach used in this work involves simulating a data stream by stepping through historical data in a sequential manner. The aim is to track key performance aspects of the predictive model as a function of time as well as also quantifying the level of drift in the features used by the model that determine build outcomes over the progression of the data stream. This experimentation revealed that the model was robust to concept drift as the overall classification accuracy recorded a steady increase over time.

The process of extracting and cleaning the data from the Jazz repository has been fully documented in previous work [7]. The goal of experimentation is to attempt to classify the likely outcome of a build based on the values of the measures derived from the calculated software metrics given in Table 1 and determine whether the model is more stable than those produced by traditional data mining techniques. In this case, the actual outcome of a build is known from the repository data which allows the accuracy of prediction to be calculated for the model after each build in the simulated stream is added. In this case, accuracy is defined as the total number of builds classified by the model such that outcome of the classification is the same as the known build outcome.

### 5.1 Hoeffding Tree Classification

The first phase of experimentation involves checking the accuracy of the data stream mining approach. This is related to the mining phase of the KDD process. The graph presented in Figure 6 presents the trend of overall classification accuracy for builds over time using the Hoeffding Tree method for the RSA after state software metrics. Figure 6 shows that the trajectory is on an upward path continuously with accuracy reaching its highest value in the last phase. This is to be expected, as more instances are used for training, the model is able to better discriminate between successful and failed builds, thus consequently increasing accuracy.

Table 2 shows the evolution of the model through its phases in terms of mean, standard deviation and 95% confidence intervals that was produced through a one-way ANOVA with accuracy as the dependent variable and phase as the factor variable. This is related to the assimilation phase of the KDD process. It is evident from Table 2 that the confidence intervals between different phases are non-overlapping, indicating that statistically significant differences between the phases exist. This was confirmed with a p value of 0.0, thus confirming significant differences (growth) in accuracy with the progression of the stream. This same trend is repeated for true positive and false positive rates for successful and failed builds, which will be presented subsequently.

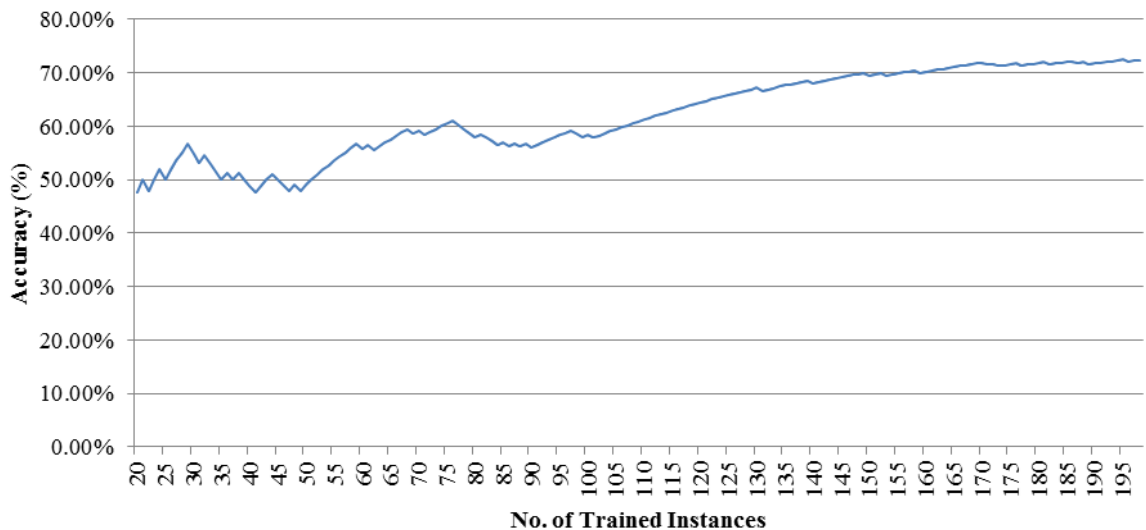

**Figure 6. Hoeffding Tree Overall Classification Accuracy**

| Phase | Instance Number | Cumulative Instance Count | Mean | Std. Deviation | Std. Error | 95% Confidence Interval for Mean | |
|---|---|---|---|---|---|---|---|
| | | | | | | *Lower Bound* | *Upper Bound* |
| 1 | 20 | 40 | .5175 | .02468 | .00552 | .5059 | .5291 |
| 2 | 40 | 80 | .5478 | .04347 | .00687 | .5338 | .5617 |
| 3 | 100 | 180 | .6520 | .05380 | .00538 | .6413 | .6627 |
| 4 | 18 | 198 | .7200 | .00000 | .00000 | .7200 | .7200 |

**Table 2. Variation of Overall Classification Accuracy by Phase**

To expand on the overall accuracy findings, Figure 7 presents the trend of classification accuracy for builds in the simulated data stream that were successful. As the actual outcome of each build was known in advance it is possible to report the outcomes of the classification in more detail in terms of true positive and false positive measures. In this context, a true positive is a classification outcome that accurately reflects the tested-for class. So a build that is known to have failed would actually be classified as a failed build by the model would be considered to be a true positive. Similarly a build that is known to be successful that is classified by the model as successful would also correspond to a true positive. A build that is known to have failed that is classified as a successful build would therefore be a false negative. These measures are presented in Figure 7 for successful builds. For the case of successful builds, the true positive rate is the proportion of successful builds that have been correctly classed as successful builds. The false positive rate is proportion of failed builds that have been incorrectly misclassified as successful builds. At the beginning of the time series in Phase the true positive rate was relatively low as shown in Figure 7. As with the overall accuracy a one-way ANOVA is performed on the true positive rate for successful builds and Table 3 shows that the 95% confidence interval in Phase 1 was [49%, 54%]. The trajectory of the classification accuracy curve showed a steady increase in accuracy with an increase in the number of training examples, attaining an accuracy of 81% towards the end of the time series in Phase 4, with a 95% confidence interval of [81%,82%], as shown in Table 3.

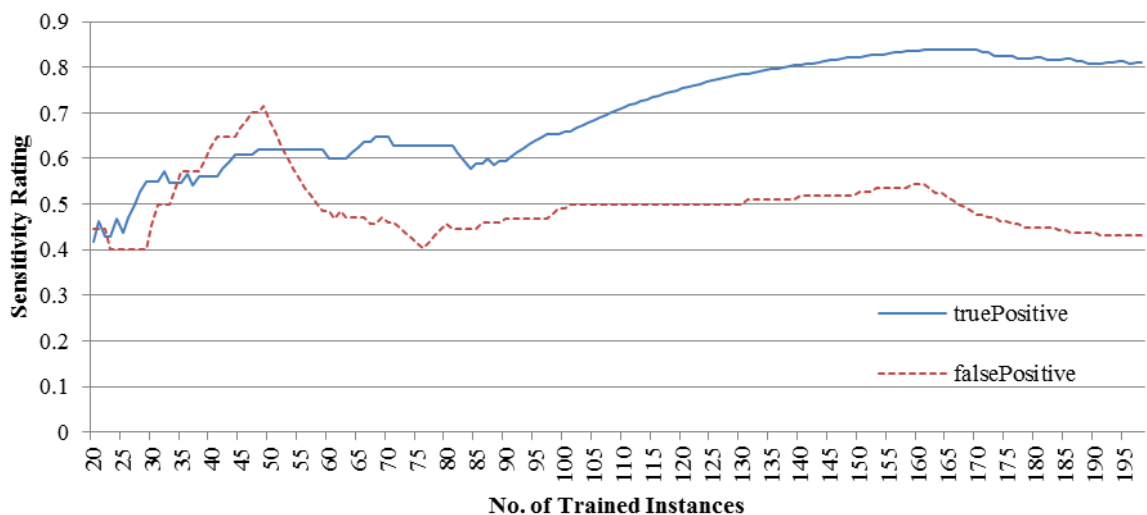

**Figure 7. Hoeffding Tree Sensitivity Measurements for Successful Builds**

| Phase | Number of instances | Cumulative Instance Count | Mean | Std. Deviation | Std. Error | 95% Confidence Interval for Mean | |
|---|---|---|---|---|---|---|---|
| | | | | | | *Lower Bound* | *Upper Bound* |
| 1 | 20 | 40 | .511065 | .0536746 | .0120020 | .485945 | .536185 |
| 2 | 40 | 80 | .616892 | .0197262 | .0031190 | .610584 | .623201 |
| 3 | 100 | 180 | .751536 | .0835939 | .0083594 | .734949 | .768123 |
| 4 | 18 | 198 | .813647 | .0048405 | .0011105 | .811314 | .815980 |

**Table 3. True Positive Rate for Successful Builds**

Table 3 also shows that the confidence intervals for the different phases are very much distinct from each other and the ANOVA analysis yielded a p-value of 0.0, showing that there exist statistically distinct differences in the true positive rates at the 95% confidence level.

The sensitivity measures indicate that there is a period of instability for correctly predicting success. As with the overall prediction accuracy, after approximately 80 builds there is a gradual but steady increase in the ability to correctly classify successful builds.

On the other hand, both Figure 6 and Table 5 show that that correctly classifying failed builds is harder to achieve. This result is consistent with previous work [31]. The percentage of failed builds classified as successful builds hovers at around 50% up until Phase 4 where a significant improvement is seen. The significance of what is occurring in terms of the model evolution at the start of Phase 4 at the 180 build mark will be analyzed in section 5.2.

| Phase | Number of instances | Cumulative Instance Count | Mean | Std. Deviation | Std. Error | 95% Confidence Interval for Mean | |
|---|---|---|---|---|---|---|---|
| | | | | | | *Lower Bound* | *Upper Bound* |
| 1 | 20 | 40 | .475590 | .0730182 | .0163274 | .441416 | .509764 |
| 2 | 40 | 80 | .536368 | .0994181 | .0157194 | .504572 | .568163 |
| 3 | 100 | 180 | .496524 | .0261319 | .0026132 | .491339 | .501709 |
| 4 | 18 | 198 | .437411 | .0074538 | .0017100 | .433818 | .441003 |

**Table 4. False Positive Rate for Successful Builds**

In Figure 8 an interesting result is observed when classifying failed build instances. In this time series after about 80 training instances the trend for false positive ratings first stabilizes and thereafter starts to decrease. In a training set with few failed instances, there are more instances that are predicted as failures incorrectly than there are instances of failed builds been predicted correctly. However, as the stream progresses and more failed signatures are seen the false positive rate drops substantially, as shown in Figure 8. This trend is reinforced in the ANOVA analysis presented in Table 6 where statistically significant drops in the false positive rates in Phases 3 and 4 of the training process are observed.

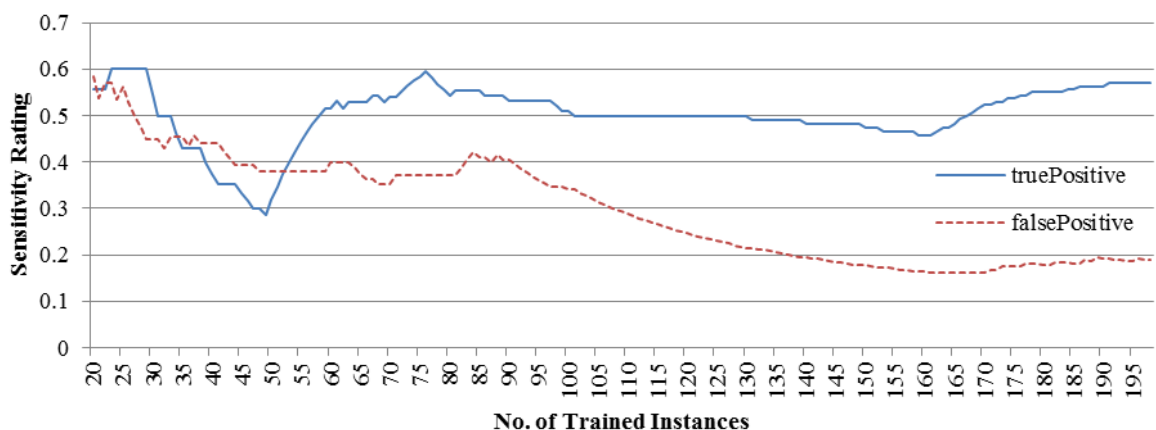

**Figure 8. Hoeffding Tree Sensitivity Measurements for Failed Builds**

| Phase | Instance Number | Cumulative Instance Count | Mean | Std. Deviation | Std. Error | 95% Confidence Interval for Mean | |
|---|---|---|---|---|---|---|---|
| | | | | | | Lower Bound | Upper Bound |
| 1 | 20 | 40 | .524410 | .0730182 | .0163274 | .490236 | .558584 |
| 2 | 40 | 80 | .463632 | .0994181 | .0157194 | .431837 | .495428 |
| 3 | 100 | 180 | .503476 | .0261319 | .0026132 | .498291 | .508661 |
| 4 | 18 | 198 | .562589 | .0074538 | .0017100 | .558997 | .566182 |

**Table 5. True Positive Rate for Failed Builds**

| Phase | Number of instances | Cumulative Instance Count | Mean | Std. Deviation | Std. Error | 95% Confidence Interval for Mean | |
|---|---|---|---|---|---|---|---|
| | | | | | | Lower Bound | Upper Bound |
| 1 | 20 | 40 | .488935 | .0536746 | .0120020 | .463815 | .514055 |
| 2 | 40 | 80 | .383107 | .0197262 | .0031190 | .376799 | .389416 |
| 3 | 100 | 180 | .248464 | .0835939 | .0083594 | .231877 | .265051 |
| 4 | 18 | 198 | .186353 | .0048405 | .0011105 | .184020 | .188686 |

**Table 6. False Positive Rate for Failed Builds**

Our experimentation with the Hoeffding tree classification method so far has concentrated exclusively on inducing a decision tree model by training with instances or builds that were in chronological order of time, from earliest to latest. The effects of varying the order in which instances are used in training the decision tree model has also been investigated. This involved generating 10 different sequences, each containing a different time ordering of build instances as described in section 4 and applied a one-way ANOVA with sequence as the factor variable. The last sequence, S10 represents pure chronological time ordering and thus represents a baseline for measuring the effect of ordering on classification accuracy. In order to eliminate the confounding effect of size of training set on accuracy, accuracy was measured over the last 21 instances, a point at which the model had benefited from receiving 177 training instances. This is the final verification of the model and is related to the assimilation phase of the KDD process.

Table 7 shows that sequence S10 had the best classification accuracy. The ANOVA reported a statistically significant difference in accuracy amongst the sequences with a p value 0.0. Having established that order of training instances is a significant factor the next step was to determine whether chronological time ordering was in fact superior. This was established by conducting two separate contrasts. In the first of the two contrasts we compared the mean from S10 with that of S9 and a p value of 0.002 was obtained, showing a statistically significant difference at the 95% confidence level. In the next contrast the mean from S10 is compared with that from S1. Again statistical significance was obtained at the 95% confidence level with a p value of 0.0. This experiment thus establishes the value of training the model with builds that are ordered over time.

| Group | Group Size | Mean | Std. Deviation | Std. Error | 95% Confidence Interval for Mean | |
|---|---|---|---|---|---|---|
| | | | | | *Lower Bound* | *Upper Bound* |
| 1 | 21 | .7090 | .01300 | .00284 | .7031 | .7150 |
| 2 | 21 | .6810 | .00301 | .00066 | .6796 | .6823 |
| 3 | 21 | .6433 | .00856 | .00187 | .6394 | .6472 |
| 4 | 21 | .6819 | .00602 | .00131 | .6792 | .6846 |
| 5 | 21 | .6648 | .00602 | .00131 | .6620 | .6675 |
| 6 | 21 | .6424 | .00944 | .00206 | .6381 | .6467 |
| 7 | 21 | .6552 | .01123 | .00245 | .6501 | .6604 |
| 8 | 21 | .6671 | .00717 | .00156 | .6639 | .6704 |
| 9 | 21 | .7114 | .00793 | .00173 | .7078 | .7150 |
| 10 | 21 | .7200 | .00000 | .00000 | .7200 | .7200 |

**Table 7.** Classification Accuracy by Sequence Order

Figure 9 presents the final decision tree generated from the stream mining process using the Hoeffding Tree method with training sequence S10. The leaves of the tree show the predicted outcome and the numeric values represent the votes used in the majority vote classifier. The value on left represents the weighted votes for failed builds and the value on the right represented the weighted votes for successful builds (i.e. failed builds | successful builds).

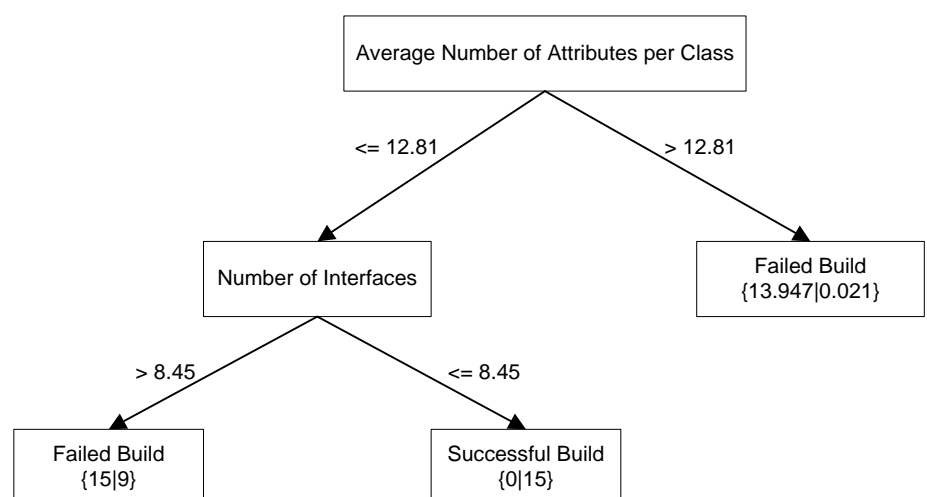

**Figure 9.** Final Hoeffding Tree for After State Software Metrics

Here it is observed that only 2 of the available features are used to classify the build outcome for all instances. At the root of the tree the first decision factor is Average Number of Attributes per Class, indicating that this measure has a high impact on the classification of results. The second decision factor is the Number of Interfaces. Figure 9 shows that failed builds are associated with a high Average Number of Attributes per Class. This is intuitive because the higher the number of attributes the more complex a class may become. If the Average Number of Attributes value per Class is low and the Number of Interfaces is high then this model also predicts a failed build. Too many interfaces have the potential lead to a design defect known as the "Swiss Army Knife" anti-pattern [54], where a complex class uses a high number of interfaces to address many different requirements. This problem may be due to rapid changing, or misinterpretation of system requirements. Interestingly, Figure 9 shows that failures are predicted with a much greater degree of confidence on the basis of the Average Number of Attributes per class feature when compared to the Number of Interfaces feature.

### 5.2 *Concept Drift Detection Results*
It is observed that through learning from the 198 instances a total of 50 concepts drifts occur. Figure 10 and Figure 11 show the concept changes that occur with respect to the two predictor features that the Hoeffding tree model uses. In both diagrams, the dotted line indicates when a concept drift is detected, irrespective of whether the change was triggered by a change in the particular measure. A step change in the cumulative drift detection indicates that a concept drift has been detected. As is evident from the figures, the concept changes are first fairly chaotic, with large values triggering changes as particular builds are added to the data stream. This response is not unexpected as the relatively small dataset makes the emerging model sensitive to extreme changes in any given measure. Over time, the concept drift rate slows down, reflecting greater stability in the data streaming in, enabling overall classification accuracy to improve without the need for structural changes in the model. This builds on the validation of the model and is the first step of the second experimental phase that addresses the goal of this research.

The subtle nature of the concept changes would indeed make it difficult, if not impossible for a human designer to decide when such changes occur without the use of an automated change detector such as ADWIN. Apart from the need to update the decision tree model when such changes occur, the detection of such changes are useful in their own right to the human user as they represent changes in patterns, arising from changes in the software development environment that would be of use to developers and project managers, alike.

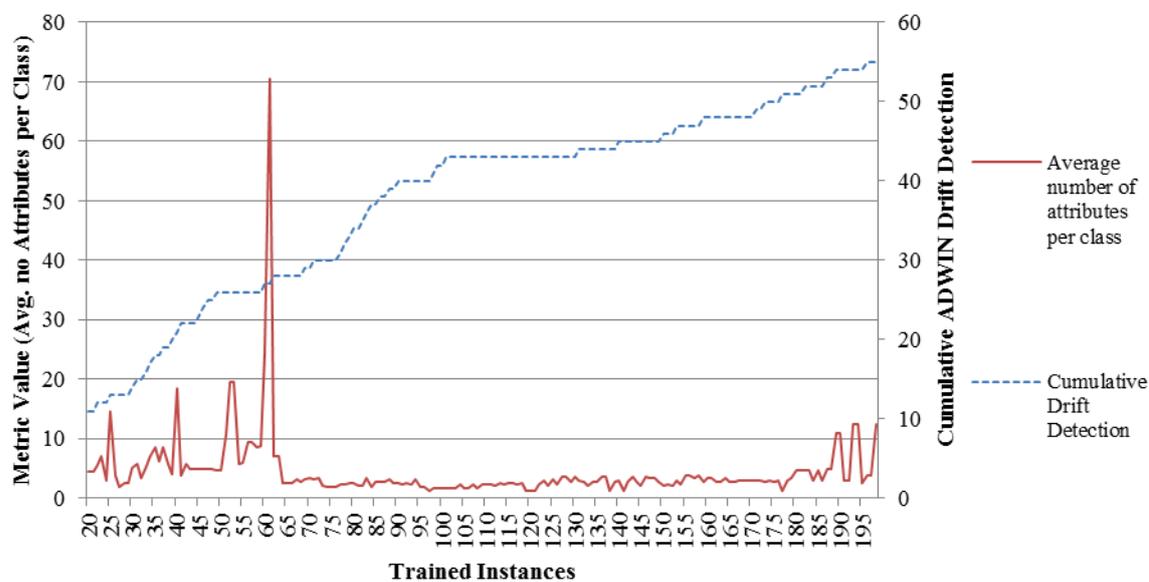

**Figure 10. Trajectories of the Average Number of Attributes per Class feature and Cumulative Drift Count over time**

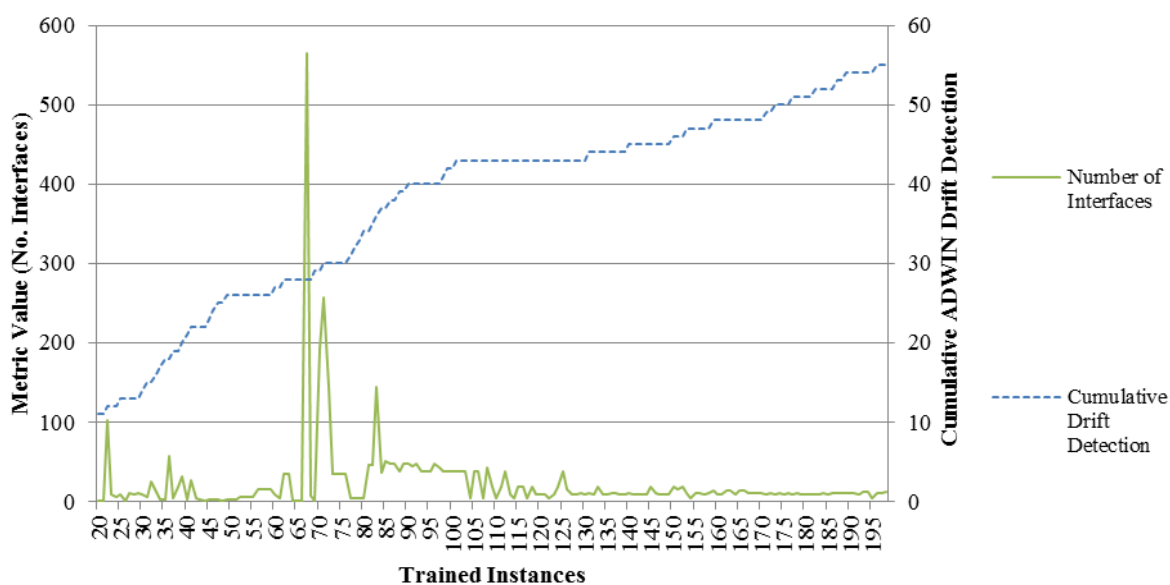

**Figure 11. Trajectories of the Number of Interfaces feature and Cumulative Drift Count over time**

*5.3 Model Comparison & Evaluation*

Applying the Hoeffding tree analysis to the Jazz data stream results in the classification model shown in Figure 9. To fully understand the application of the Hoeffding tree approach and evaluate the stability of the models it is important to analyze the emergence of this model, not just the final model itself. By examining Figure 4 it would seem reasonable to conclude that the minimum number of instances required to develop a classification tree that is reasonably stable would be around 100 instances. It is at this point that the prediction accuracy starts to stabilize and show a trend to improving asymptotically. However, an examination of the Hoeffding tree analysis at this point shows that no actual decision tree has been generated by the approach. In fact, the Hoeffding tree approach has not identified a single feature that has sufficient predictive power to use effectively. The approach is therefore attempting to classify a new build in the data stream against the majority taken over all instances and all attribute values. So, for example, if there were 60% successful builds, then all builds would be labeled success. This is an exceptionally degenerate case where severe model under-fitting is occurring due to lack of training examples resulting in no clear predictors. The Hoeffding tree approach identifies an actual decision tree only after 160 builds. This first decision tree identifies only a single attribute against which to classify a given build and the resulting decision tree is shown in Figure 12.

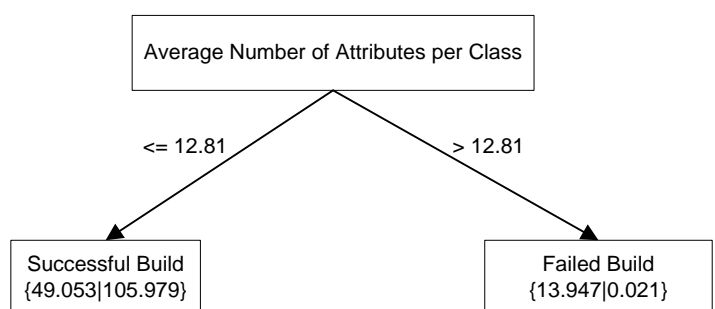

**Figure 12. Initial Hoeffding Tree Model**

The emergence of this initial model has an impact on the ability of the Hoeffding tree approach in terms of classifying build outcomes. In Figure 7 it can be observed that there is a drop in the number of failed builds being misclassified as successful builds that is a direct result of the emergence of the model.

The final model in Figure 9 emerges after 180 builds, which is an indication that sufficient new data has emerged to enable the Hoeffding tree analysis to identify a statistically significant change in the data stream through a number of concept drift detections. The detection of statistically significant changes is effected by ADWIN. The ADWIN change detector uses a window of examples (which comprises all instance from the last change detection point up to the current instance) and tests every possible partition of any given window into two sub-windows, $W_a$ and $W_r$. In each sub-window the mean error rate of the classification model is computed. A significant change in the data triggers a spike in the error rate and a statistical test on the means of the two sub-windows can be used to detect such changes. In the case of ADWIN, the empirical Bernstein bound is bound is used as the basis of the statistical test which is conducted by ADWIN at its default confidence level of 99%.

The impact of this change is apparent in Figure 8 where an incremental improvement in classification of failed builds is apparent. The ability to react to only statistically significant changes is one of the key advantages of the approach over simple classification methods such as the J48 algorithm, which do not attempt to detect statistical significance and as a result are prone to very large changes in the resulting decision trees as a result of very small changes in the underlying data. This phenomenon has been observed in previous work [6, 7] where it was observed that very small changes to the underlying data could produce very large differences in the models produced by techniques such as the J48 decision tree or a Naïve Bayes classifier.

To some extent, this observation can be further confirmed by applying such traditional data mining approaches to subsets of the data that represent different snapshots of the data stream. Figure 13 is the decision tree that results from applying the J48 algorithm to the first 160 build instances of the data stream whilst using 10-folds cross-validation. This is the same number of builds as the Hoeffding tree approach uses to create the first decision tree model shown in Figure 12.

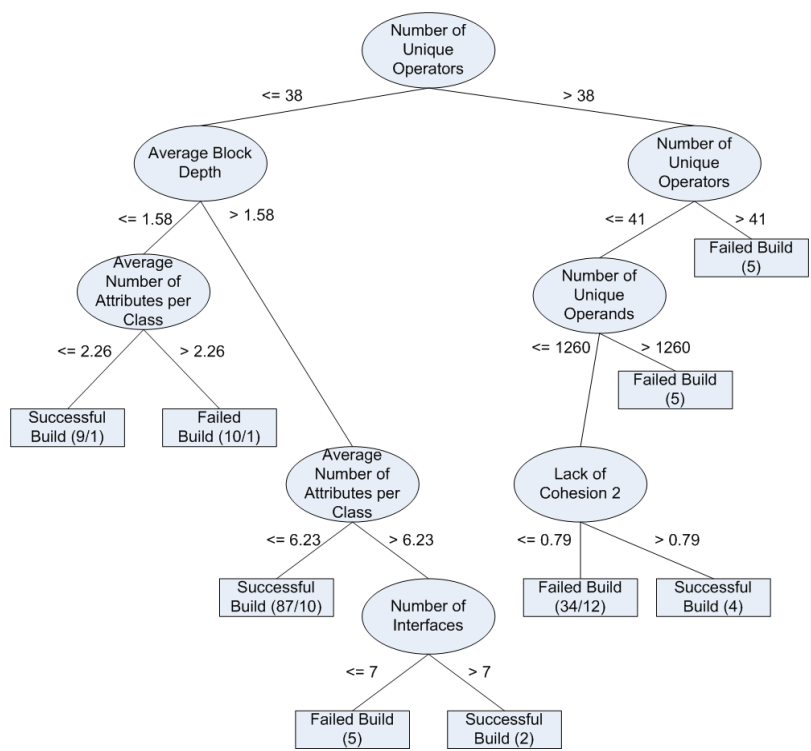

**Figure 13. J48 Decision Tree (160 builds)**

The first observation to make regarding this decision tree appears much more complex nature than that which emerges from applying the Hoeffding tree analysis. Bearing in the mind the relative optimism associated with cross-validation, in this case the overall accuracy of the model produced by the J48 algorithm is 71.9% and the true positive rate for failed builds is 0.544. The overall accuracy is comparable to that of the Hoeffding tree model but the true positive rate is marginally higher. It is worth noting that some of the features selected by the Hoeffding tree approach are present in the decision tree produced by applying the J48 algorithm; for example a very large number of the build instances are classified by "Average Number of Attributes per Class", though there is a significant difference in the threshold value used for the classification.

Figure 14 shows the decision tree obtained by applying the J48 algorithm to the first 180 builds in the data stream, which corresponds to the number of builds where the Hoeffding tree approach evolves the classification tree. The decision tree has evolved because sufficient numbers of concept drifts have been detected to reinforce the change in pattern observed in the data stream.

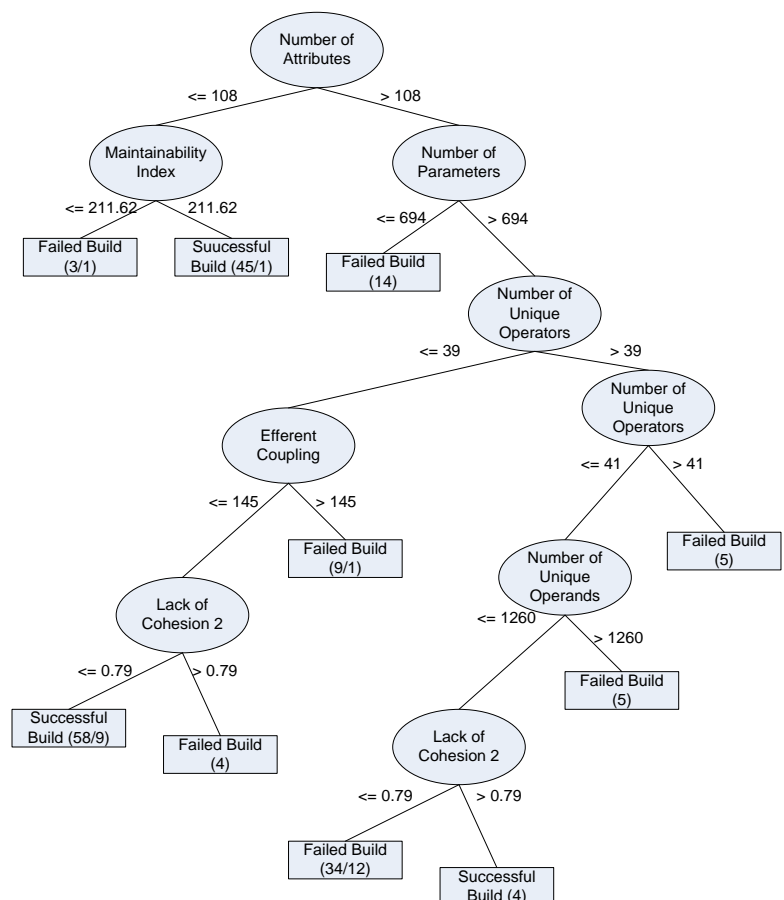

**Figure 14. J48 Decision Tree (180 builds)**

In this case the first observation is that this decision tree is radically different to that presented in Figure 13, despite there being only a relatively small number of additional builds. This lack of stability has been noted in previous work [7] and the goal of this work is to determine whether the Hoeffding Tree approach produces more stable models than traditional approaches such as the J48 algorithm. In order to show this is the case, the final decision trees for each approach are compared to the preceding tree using a number of measures, including the attribute churn measure defined in section 4. These measures are shown in Table 8.

|  | J48 (160 builds) | J48 (180 builds) | Hoeffding Tree (160 builds) | Hoeffding Tree (180 builds) |
| --- | --- | --- | --- | --- |
| Size (depth of tree) | 4 | 6 | 1 | 2 |
| Size (no. tests needed) | 8 | 9 | 1 | 2 |
| Size (no. of leaf nodes) | 9 | 10 | 2 | 3 |
| No. Attributes (total) | 6 | 7 | 1 | 2 |
| Attribute Churn | - | 100 | - | 50 |

**Table 8.** Comparison of Models

In general, the J48 algorithm could not be applied in a data stream environment since it requires that all data be available at one point in time to build the model. However, in the software engineering context where the build cycle is measured in weeks, the model could be rebuilt following each build attempt. Despite the slightly higher accuracies, which potentially are attributed to the use of cross-folds validation, the use of J48 has some drawbacks. The J48 algorithm has responded to changes without any consideration of whether the new builds introduce a statistically significant alteration from previous history. The J48 algorithm responds whether or not there is a consistent pattern of change in the data stream and as a result is demonstrably less stable. The overall classification accuracy of this decision tree is 71.7% and the true positive rate of failed builds is 0.652 which again is marginally higher than that achieved through the application of the Hoeffding tree model, though it is worth noting that the use of 10-folds cross-validation does potentially result in a much more optimistic model so this difference may in practice not be present. The purpose of comparing the outcomes is to show the value of using the ADWIN approach for concept drift detection and the result in terms of the Hoeffding tree approach, that results in only detecting and responding to consistent patterns of change. A more stable model that exhibits incremental change is more likely to be of use to a software development team.

In this case, none of the features present in the decision tree developed by applying J48 are the same as those in the Hoeffding tree model. It is worth noting that there are some similar features, though. For example the presence of Number of Attributes as opposed to the Average Number of Attributes per class. One of the challenges in dealing with software metrics has been the fact that many metrics are in fact derived from other metrics which makes it more difficult to identify the truly significant features. This also contributes to the difference observed between the two models produced by applying J48. Applying J48 would result in a highly inconsistent set of decision trees over time that would be unlikely to stabilize to the selection of a meaningful and consistent set of features until vastly more data is available. Comparatively, the Hoeffding tree approach produces a smaller model that evolves, as opposed to changes, when statistically significant changes occur. This produces a decision tool much more useful to software development teams.

However, it is clear that the number of builds in the data set is still not sufficient to build a robust and stable predictive model though an analysis of Figure 8 shows that the emergence of the initial decision tree at 160 builds has the potential to be the start of a much more useful model. The introduction of the decision tree increases the capability of the model to identify failed builds, as apparent from the increase in true positive rate as well as an increase in both precision and recall measures. It is likely that with more data that the asymptotic nature of the graph in Figure 10 would alter when more significant predictors emerge and the overall ability to correctly classify both failed and successful builds would increase.

## 6. Limitations & future work

Most of the limitations in the current study are products of the relatively small sample size of build data from the Jazz project combined with the sparseness of the data itself. For example, the ratio of metrics (48) to builds (199) is such that it is difficult to truly identify significant metrics. In particular, it has been observed that predicting failure is more challenging than predicting success and that not predicting failure doesn't mean that success has been predicted. This is due to the fact that the build successes and failures overlap in feature space and "failure" signatures have a greater degree of fragmentation than their "success" counterparts. This overlap is a strong symptom of the fact that some vital predictors of software build failure have not been captured in the Jazz repository. It is an open question as to whether such predictors can indeed be quantified in a form suitable for use in a machine learning predictive context.

As such, one aspect of future work is to develop a deeper understanding of what source code characteristics are most related to build failure and attempt to develop a set of indicative metrics that can provide development teams with the opportunity to proactively manage risk exposure throughout a development which is just as important, if not more important than unequivocally predicting build failure or success.

## 7. Conclusions

This paper presents the outcomes of an initial systematic attempt to predict build success and/or failure for a software product by utilizing source code metrics. Overall prediction accuracies of 72% have been achieved through the use of the Hoeffding Tree algorithm, a data stream mining technique. The data stream mining approach has been shown to produce a model that evolves incrementally over time and as such is more stable than other classification tree approaches.

Despite this high overall accuracy, there is greater difficulty in predicting failure than success and at present the classification trees content some uncertainty and confusion, but show promise in terms of informing software development activities in order to minimize the chance of failure. This result is similar to previous work, where traditional data mining methods were used to explored the Jazz dataset [6]. However, the application of datastream mining techniques has produced more stable and robust models that are more likely to be adopted by a development team.

This research has presented a potential solution for encoding software metrics as data streams. In the case of Jazz the data streams would be provided when a software build was executed, though this study simulated such a data stream from historical data. The real-time streams can be run against the model which has been generated from software build histories. From the real-time based predictions developers may delay a build to proactively make changes on a failed build prediction. One of the advantages of building predictive models using data stream mining methods is that they do not have large permanent storage requirements. The main reason why Jazz only stores a limited number of build change sets is because of the huge storage requirements.

The results have shown that data stream mining techniques holds much potential as the Jazz environment, as the platform, can continue to store the latest *n* builds without losing relevant information for a prediction model that has been built over an extended series of (older) software builds. As a tool, the predictive models can be encoded into the IDE and updated when builds are performed. This tool would provide contributors with real-time feedback during the development of their code in relation to the metrics extracted and predicted build outcome. It would also provide real-time insights into the way the team is communicating effectively for generating a successful build. While data stream mining has application in managing network, web searches traffic, systems, networks/data, ATM transactions and safety, few studies have studied data stream mining using software metrics. To our knowledge this is the first attempt ever made to use data stream mining techniques for predicting software build outcomes using software source code metrics.

## 8. Acknowledgements

Our thanks go to IBM for providing access to the Jazz repository and the BuildIT consortium that has funded this research. We also like to thank Professor Stephen MacDonell from Auckland University of Technology for providing valuable expertise regarding software metrics.